\documentclass[twocolumn,pra,aps]{revtex4}
\usepackage{graphicx}
\usepackage{amsmath}

\begin{document}

\title{Femtosecond Time-Bin Entangled Qubits for Quantum Communication}
\author{I. Marcikic$^\dag$, H. de Riedmatten$^\dag$, W. Tittel$^{\dag,\ddag}$, V. Scarani$^\dag$, H. Zbinden$^\dag$ and N.
Gisin$^\dag$}

\affiliation{$^\dag$Group of Applied Physics-Optique, University of
Geneva, CH-1211, Geneva 4,
Switzerland\\
$^\ddag$Danish Quantum Optics  Center, Institute for Physics and Astronomy, University of Aarhus,
Denmark}

\begin{abstract}
We create pairs of non-degenerate time-bin entangled photons at telecom
wavelengths with ultra-short pump pulses. Entanglement is shown by performing
 Bell kind tests of the Franson type with visibilities of up to 91\,\%. As
time-bin entanglement can easily be protected from decoherence as
encountered in optical fibers, this experiment opens the road for
complex quantum communication protocols over long distances. We also investigate the creation
of more than one photon pair in a laser pulse and present a simple tool to quantify
the probability of such events to happen.
\end{abstract}

\maketitle

\section{ Introduction}

Entanglement is one of the most important tools for the realization of complex
quantum communication protocols, like quantum teleportation or entanglement
swapping, and due to their ability to be transported in optical fibers,
photons are the best candidates for long distance applications \cite{weihs}.
Even though some of these protocols have already been experimentally
realized \cite{dik,Boschi,fur,kim,martini,Pan,thomas}, none of them was
optimized for long distance communication. Most of them used polarization entangled
photon pairs in the visible range which are subject to important
attenuation, and suffer from decoherence (depolarization) due to
polarization mode dispersion (birefringence) in optical fibers. Energy-time
entanglement or its discrete version, time-bin entanglement \cite{jurgen},
is more robust for long distance applications. Both types have been proven
to be well suited for transmission over more than 10 km \cite{10km,rob}, and
have already been used for quantum cryptography \cite
{crypto1,crypto2}. However these experiments did not rely on joint
measurements of photons from different pairs where the emission time of each
pair must be defined to much higher precision. For this purpose we built and
tested a new source using femtosecond pump pulses. This is the
first femtosecond source at telecommunication wavelengths, and the first
femtosecond source employing time-bin entanglement. This will allow
realization of teleportation and entanglement swapping over long distances.

Apart from ensuring good localization of the photon pairs a
femtosecond pulse engenders a significant probability of
creating a pair per pulse due to the high energy contained in each
pulse, an important requirement when two pairs have to
be created at the same time. However when this probability becomes
significant, the probability of creating unwanted multiple pairs becomes higher.
Thus, the purity of entanglement will decrease, a phenomenon which is
unwanted for almost all quantum communication protocols (Bell test, cryptography,
teleportation etc.). For instance, the photon pair visibility in a
Bell type test will strongly depend on the relation between the multiple pairs.
They can be either independent or they can be described as multiphoton entanglement.

In the following we first remind the reader of the basic principle of
time-bin entanglement, and we explain how to test entanglement. We
then describe the experimental setup we used and present the
results. In addition, we experimentally verify the reduction of
the visibility due to multiple pair creation. Finally, we present a
straightforward measurement of the probability to create a pair per
pulse.

\section{Femtosecond time-bin entanglement}

A time-bin qubit is formed by a coherent superposition of
amplitudes describing a photon to be in two time-bins separated by
a time difference which is much larger than the coherence time of
the photon. It is created by a short pulse, in our case a
femtosecond pulse, passing through an unbalanced interferometer,
referred to as the pump interferometer, with a relative phase
$\varphi $ between the two arms. The output state of the photon,
after the pump interferometer, can be written as:
\begin{equation}
\left| \Psi \right\rangle _{p}=\frac{1}{\sqrt{2}}(\left| 1,0\right\rangle
-e^{i\varphi}\left| 0,1\right\rangle)  \label{1}
\end{equation}
The state $\left| n_0,n_1,n_2,...\right\rangle$ corresponds to
the case where $n_0$ photons are in the first time-bin (passing zero times through
the long arm of any interferometer), $n_1$ photons are in the second
time-bin\ (passing once through a long arm of any interferometer), $n_2$ photons are in the
third time-bin (passing through the long arms of two different interferometers) etc.
Entangled time-bin qubits are created by passing a time-bin
qubit through a non linear crystal where eventually twin photons can be
created by spontaneous parametric down conversion. The creation time is then
given by superposition of two values:
\begin{equation}
\left| \Phi \right\rangle =\frac{1}{\sqrt{2}}(\left| 1,0\right\rangle
_{A}\left| 1,0\right\rangle _{B}-e^{i\varphi }\left| 0,1\right\rangle
_{A}\left| 0,1\right\rangle _{B})  \label{2}
\end{equation}
Depending of the relative phase $\varphi ,$ two out of four Bell states can
be created ($\Phi ^{\pm }$)$.$ The two remaining Bell states ($\Psi ^{\pm }$%
) can be created in principle with switches and delays after the
crystal.

\subsection{Bell test}

\begin{figure}[h]
\includegraphics[width=8.43cm]{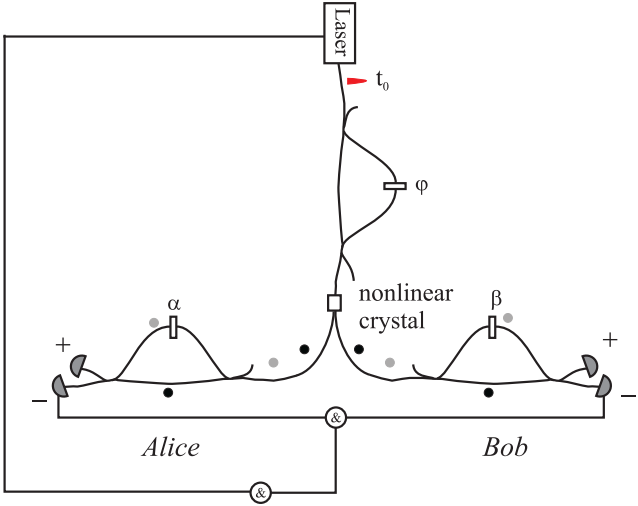}
\caption{{Scheme of a Bell type experiment using time-bin
entangled photons. Time bin qubits are prepared by passing a fs
pulse through the pump interferometer. Eventually, a pair of
entangled photons is created in the crystal. Alice and Bob analyze
the photons using interferometers that are equally unbalanced with
respect to the pump interferometer, thereby sending the amplitude
in the first (gray) time bin through the long arm and the one in
the second (black) time bin through the short arm and thus undoing
the transformation of the pump interferometer (in 50\,\% of the
cases).}}
\label{setup}
\end{figure}

\begin{figure}[h]
\includegraphics[width=8.43cm]{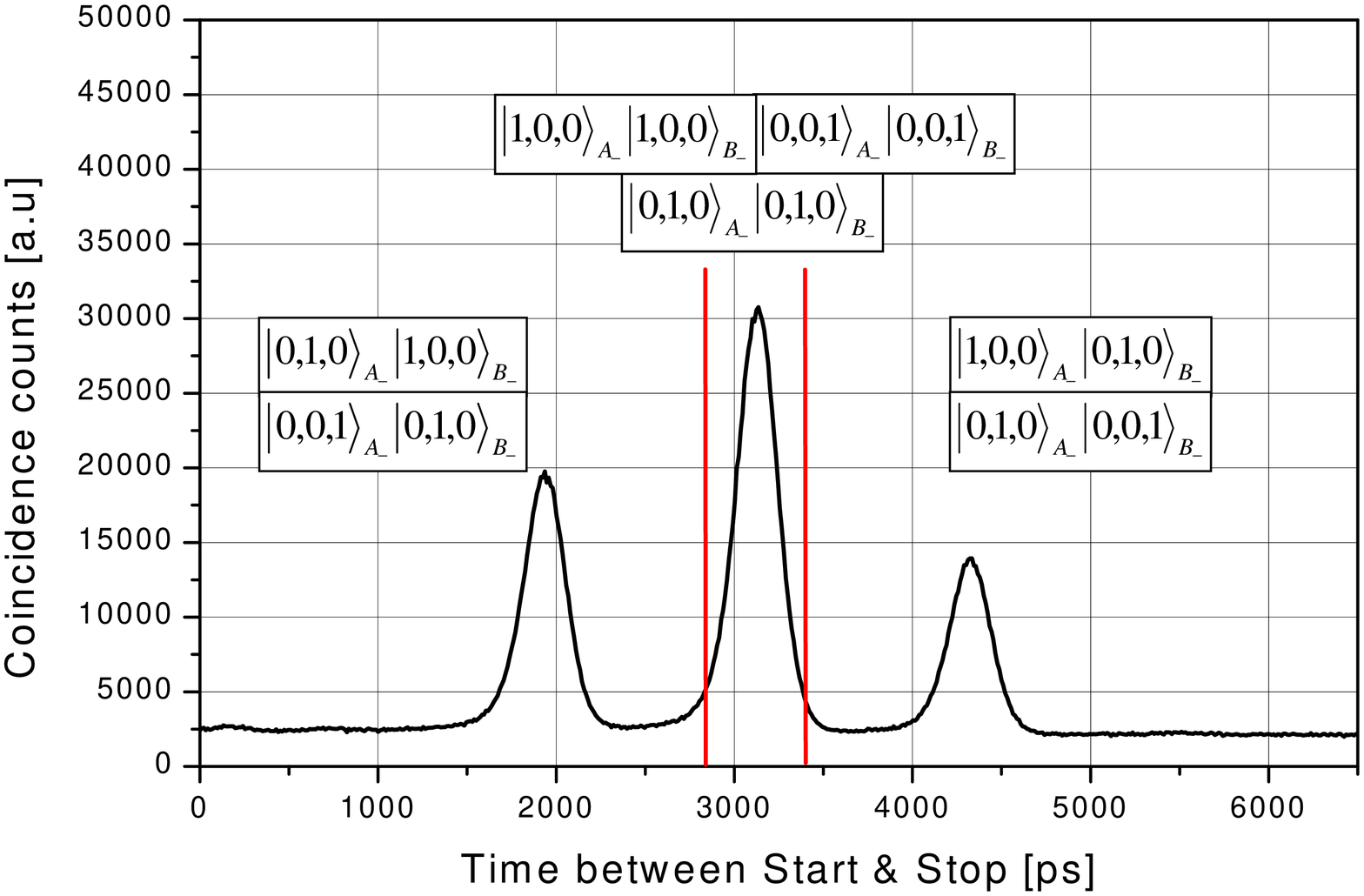}
\caption{{Time histogram of the difference of arrival times between Alice's
and Bob's detector. The spacing between two peaks corresponds to the optical
path difference in any interferometer.}}
\label{TAC}
\end{figure}

To qualify the purity and degree of entanglement we perform a Bell test
(Franson type) \cite{Franson}. One of the photons is sent to Alice and the
other one to Bob (see figure \ref{setup}).
To analyze the received qubit, Alice and Bob undo the initial transformation
with an interferometer which has the same optical path length difference as the
pump interferometer. The initial state $\left| 1,0\right\rangle_{A}$
evolves as follows:
\begin{eqnarray}
&&\hspace{-1cm}\left| 1,0\right\rangle_{A} \mapsto \nonumber \\
&&\hspace{-1cm} \frac{1}{2}[\,\left|1,0,0\right\rangle_{A_{-}}\left| 0,0,0\right\rangle_{A_{+}}-e^{i\alpha}\left|0,1,0\right\rangle_{A_{-}}\left| 0,0,0\right\rangle_{A_{+}}\nonumber \\
&&\hspace{-1cm}+\,i\left|0,0,0\right\rangle_{A_{-}}\left| 1,0,0\right\rangle_{A_{+}}
+\,ie^{i\alpha}\left|0,0,0\right\rangle_{A_{-}}\left| 0,1,0\right\rangle_{A_{+}}]
\label{3}
\end{eqnarray}
With this evolution Eq.\ref{2} becomes:
\begin{eqnarray}
\left| \Psi\right\rangle= \nonumber
\frac{1}{4\sqrt{2}}[&(e^{i(\alpha+\beta)}-e^{i\varphi})&\left|0,1,0\right\rangle_{A_{-}}\left|0,1,0\right\rangle_{B_{-}}\\
\nonumber
-&i(e^{i(\alpha+\beta)}+e^{i\varphi})&\left|0,1,0\right\rangle_{A_{-}}\left|0,1,0\right\rangle_{B_{+}}\\
\nonumber
-&i(e^{i(\alpha+\beta)}+e^{i\varphi})&\left|0,1,0\right\rangle_{A_{+}}\left|0,1,0\right\rangle_{B_{-}}\\
\nonumber
-&(e^{i(\alpha+\beta)}-e^{i\varphi})&\left|0,1,0\right\rangle_{A_{+}}\left|0,1,0\right\rangle_{B_{+}}\\
+&&\textrm{24 other terms}\hspace{.75cm}]
\end{eqnarray}
In the following discussion we are interested only in coincidences between $A_{-}$ and $B_{-}$ detectors (see figure \ref{setup}).
If we monitor the difference of arrival times of two entangled photons at Alice's and Bob's
side ($t_{A_{-}}-t_{B_{-}}$), with a time to amplitude converter (TAC) we
distinguish three different peaks (see figure \ref{TAC}).
The two satellite peaks correspond to events that are well distinguishable $%
\left| 0,1,0\right\rangle _{A_{-}}\left| 1,0,0\right\rangle _{B_{-}}$ or $\left| 0,0,1\right\rangle _{A{-}}\left|
0,1,0\right\rangle _{B_{-}}$ for the left satellite
peak and $\left| 1,0,0\right\rangle _{A_{-}}\left| 0,1,0\right\rangle _{B_{-}}$ or $\left| 0,1,0\right\rangle _{A_{-}}\left|
0,0,1\right\rangle _{B_{-}}$ for the right satellite
peak. These peaks can be discarded by selecting a sufficiently small time
window around the central peak. In the central peak three events
(due to Alice's and Bob's photons taking the same path in the respective
interferometers) are counted:
$\left| 1,0,0\right\rangle _{A_{-}}\left| 1,0,0\right\rangle _{B_{-}}$,
$\left| 0,0,1\right\rangle _{A_{-}}\left| 0,0,1\right\rangle _{B_{-}}$ and
$\left| 0,1,0\right\rangle_{A_{-}}\left| 0,1,0\right\rangle _{B_{-}}$. The first (second) event corresponds to the case when
the photons created in the first (second) time-bin pass through the short (long) arm of Alice's and Bob's interferometer.
The third event corresponds either to the case when the photons
created in the first time-bin pass through the long arm of Alice's
(acquiring a relative phase $\alpha$) and Bob's (acquiring a relative phase
$\beta$) interferometer or to the case when the photons created in the second time-bin (with a relative phase $\varphi$)
pass through the short arm of Alice's and Bob's interferometer. The impossibility to
distinguish, even in principle, via which path the photons have passed leads to interference.
Knowing the emission time of the pump pulse we can distinguish two
out of three events ($\left| 1,0,0\right\rangle _{A_{-}}\left| 1,0,0\right\rangle _{B_{-}}$ and
$\left| 0,0,1\right\rangle _{A_{-}}\left| 0,0,1\right\rangle _{B_{-}}$) thus the visibility as
observed in the two photon interference while changing the phase in one of
the three interferometers is limited to 50\,\%. To increase the visibility to
100\,\% we postselect the third event by making a three fold
coincidence between the emission time of the pump photon, and Alice's and Bob's
detection\ (see figure \ref{setup}). Thus the post selected state is:
\begin{eqnarray}
\left| \Psi \right\rangle_{postselected} = \left| 0,1,0\right\rangle_{A_{-}}\left| 0,1,0\right\rangle
_{B_{-}}\label{postselection}
\end{eqnarray}
with the amplitude of probability to be
detected
\begin{eqnarray}
A\simeq e^{i(\alpha+\beta-\varphi)}-1\nonumber
\end{eqnarray}
Where $\varphi$, $\alpha$ and $\beta$ are the relative phases of
the pump, Alice's and Bob's interferometer, respectively. The triple coincidence
counting rate is, thus, given by:
\begin{eqnarray}
R_{c}\sim 1-V\cos (\alpha +\beta -\varphi )  \label{4}
\end{eqnarray}
where V is the visibility which can in principle reach the value of 1.
We take it as the figure of merit to quantify the
entanglement. Note that correlation described by such coincidence
functions with a visibility higher than 70.7 \% can not be
described by local theories \cite{CHSH}.

\subsection{Experimental setup}

A mode locked Ti:Sapphire laser (Coherent Mira 900) produces pulses at $\lambda _{p}$=710\,nm with
150\,fs pulse width and 76\,MHz repetition rate. To remove all unwanted
infrared light the light passes through a series of dichroic mirrors,
reflecting only wavelengths centered around 710\,nm. The superposition of
discrete times is made by a bulk Michelson interferometer with a path-length
difference of 1.2\,ns \cite{foot2}. The entangled non-degenerate colinear
photons at 1310 and 1550\,nm (telecom wavelengths) are created in a KNbO$_{3}$
type I nonlinear crystal. The pump light is removed with a RG\ 1000 filter,
the twin photons are collimated into an optical fiber and separated by a
wavelength-division-multiplexer (WDM). The analyzers are two Michelson fiber
interferometers with Faraday rotator mirrors. The role of these mirrors is
to compensate any difference of polarization transformation in the two arms
of the interferometer \cite{plug,pluga}. The phase is tuned by varying the
temperature of the interferometer.

At Alice's side the photon counter at 1310\,nm is a passively quenched
Germanium APD, cooled with liquid nitrogen and working in reversed mode
above the breakdown voltage (so called Geiger mode). The quantum efficiency
is around 10\,\% for a dark count rate of 20\,kHz. At Bob's side the
photons at 1550\,nm are detected by a InGaAs APD, Peltier cooled to around
-50\,$^{\circ}$C. To obtain a good signal to noise ratio, these APDs\ have to be used in
so called gated mode. They are then operational only during a short period
(around 50\,ns) when a photon is expected to arrive. Thus, the InGaAs
APD is triggered by the Ge APD. Its quantum efficiency is around 30\,\%
for a darkcount probability of $\sim$ 10$^{-4}$\,per ns \cite{JDG}.

The twin photons, due to our phasematching conditions, have a large spectral
bandwidth of around 90\,nm. To reduce the effect of chromatic dispersion in
our interferometers we limit the spectral width of the downconverted photons
with an interference filter at Alice's side ($\Delta \lambda=40$\,nm)
\cite{foot21} and we use dispersion shifted fibers for Bob's interferometer. In addition, spectral
filtering of the 1310\,nm photons leads to a decrease of the count rate of the
Ge-detector, hence to a decrease of the trigger rate for the InGaAs APD which
enables to operate them at a higher quantum efficiency.

\subsection{Results of the measurement}

\begin{figure}[h]
\includegraphics[width=8.43cm]{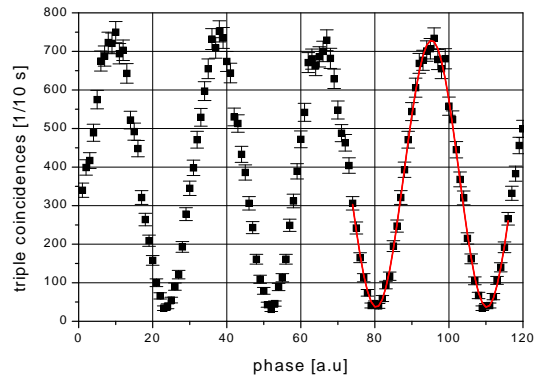}
\caption{{Net interference fringes of the triple coincidence
detection of the postselected state (Eq.\ref{postselection}).}}
\label{frange}
\end{figure}

Figure \ref{frange}\ shows the results of a typical interference curve.
The visibility of the interference fringes, after subtraction of
the noise, is 91 $\pm $ 0.8\,\% (computed using a sinusoidal fit).
This result shows that the created state is not far from a pure
maximally entangled state, sufficiently entangled to be used in
quantum communication protocols. Please note that only the net
visibility is important in this context. Indeed we have to
subtract the accidental coincidences from the raw visibility since
they are due to a combination of fiber losses, non-perfect quantum
efficiency and detector noise, and not to reduced entanglement.
However, if we assume (in addition to \cite{foot3}) that the
accidental coincidences are measured in a fair way, our net
visibility is high enough to violate the CHSH inequality
\cite{CHSH} by more than 25 standard deviations.

Note that with this source, creating entangled photons with the
same polarization and using time-bin entanglement, we did not have
problems met by other groups creating polarization entangled
photons with a femtosecond pulsed laser \cite{shih}. The quality of our
entanglement is not degraded by the use of the long crystal ($\Delta l=10$\,mm) and large
interference filters ($\Delta\lambda=40$\,nm).

\section{ Multiphoton states}

\begin{figure}[h]
\includegraphics[width=8.43cm]{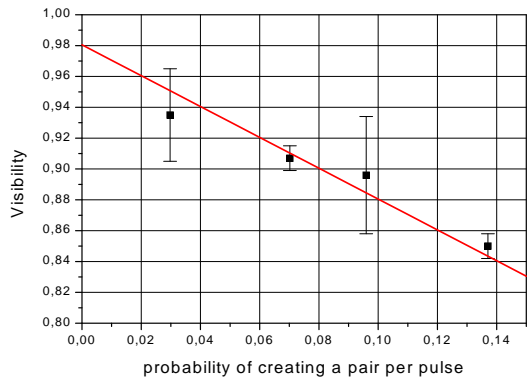}
\caption{{Decrease of the visibility as function of the pump power.
The solid line represents the theoretical predictions $V(P_{pair})=\,V_{max}-P_{pair}$, the fit yielding $V_{max}=98$\,\%}}
\label{vis}
\end{figure}

The above mentioned results were obtained using a mean pump power of 24 mW. By
increasing the pump power the probability of creating more than one pair per pulse
increases too, thus the visibility of the two-photon interference fringes
decreases. Although the pump power was chosen in order to get good
visibilities this effect is still present. Figure \ref{vis} shows the
decrease of the visibility as a function of the pump power.
The decrease of the visibility can be understood with the following simple
calculation, that can be rederived using the full formalism of quantum
optics \cite{sca}. The detection rate is the sum of two mutually incoherent
contributions: $R_2$, the detection rate associated to the production of one
pair; and $R_4$, associated to the production of four photons. The
two-photon contribution has 100\% visibility, thence we can write
\begin{eqnarray}
R_2 &=& P_{pair}\,\frac{1+\cos\theta}{2}
\end{eqnarray}
where $P_{pair}$ is the probability of creating one pair and
$\theta=\alpha+\beta-\varphi$. We discuss the four-photon contribution
supposing that the four-photon state is actually two independent pairs,
which is not strictly speaking true, but is a good guide for the
intuition --- moreover, the final result turns out to be independent of this
assumption \cite{sca}. Thus we have two possible cases: when the two photons
that are detected belong to the same pair, $R_4$ shows full interference;
when they belong to different pairs, $R_4$ shows no interference at all.
Each of the situations happens twice, because the two pairs may have been
created either both in the same pulse, or one in each pulse. Thus
\begin{eqnarray}
R_4 &=& P_{4photons}\,\left(2\,\frac{1+\cos\theta}{2}+2\,\frac{1}{2}\right)\nonumber \\
&=& \,4\,P_{4photons}\, \frac{1+\frac{1}{2}\cos\theta}{2}
\end{eqnarray}
Now assuming a Poissonian distribution for counting of independent events,
the probability of creating four photons is $P_{4photons}=\frac{P_{pair}^2}{2}$. So
finally
\begin{eqnarray}
R_c &=& \frac{1}{2}[(P_{pair}+2P_{pair}^2)+ (P_{pair}+P_{pair}^2)\cos\theta]
\end{eqnarray}
whence the total visibility is $V=\frac{1+P_{pair}}{1+2P_{pair}}\approx 1-P_{pair}$,
predicting a slope of $-1$.
\section{Characterization of the source}

As we have seen in the last section it is important to get a fast and
reliable estimation of the probability of creating a pair per pulse. Usually this
probability is computed from:
\begin{equation}
P_{pair}=N(singles)/t_{A}\eta _{A}f  \label{standard}
\end{equation}
where $P_{pair}$ is the probability of creating a pair per pulse, $%
N(singles) $ is the number of photons detected by Alice, $t_{A}$ characterizes coupling and transmission, $\eta _{A}$
is the quantum efficiency of Alice's detector and $f$ is the laser frequency. In
this case we have to estimate the values of $t_{A}$ and $\eta _{A}$ (the
quantum efficiency can be measured but it is not a straightforward
measurement).

\begin{figure}[h]
\includegraphics[width=8.43cm]{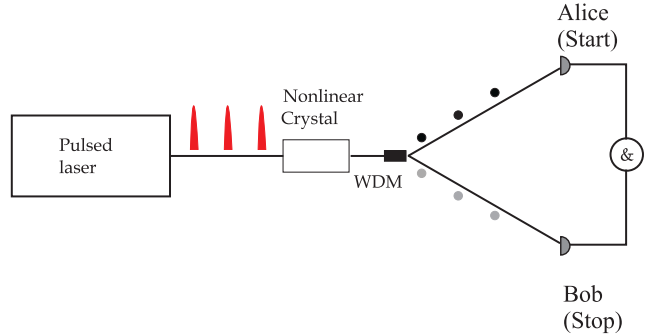}
\caption{{Experimental setup for the measurement of the probability of
creation of a pair per pulse}}
\label{set side}
\end{figure}

\begin{figure}[h]
\includegraphics[width=8.43cm, height=5.1cm]{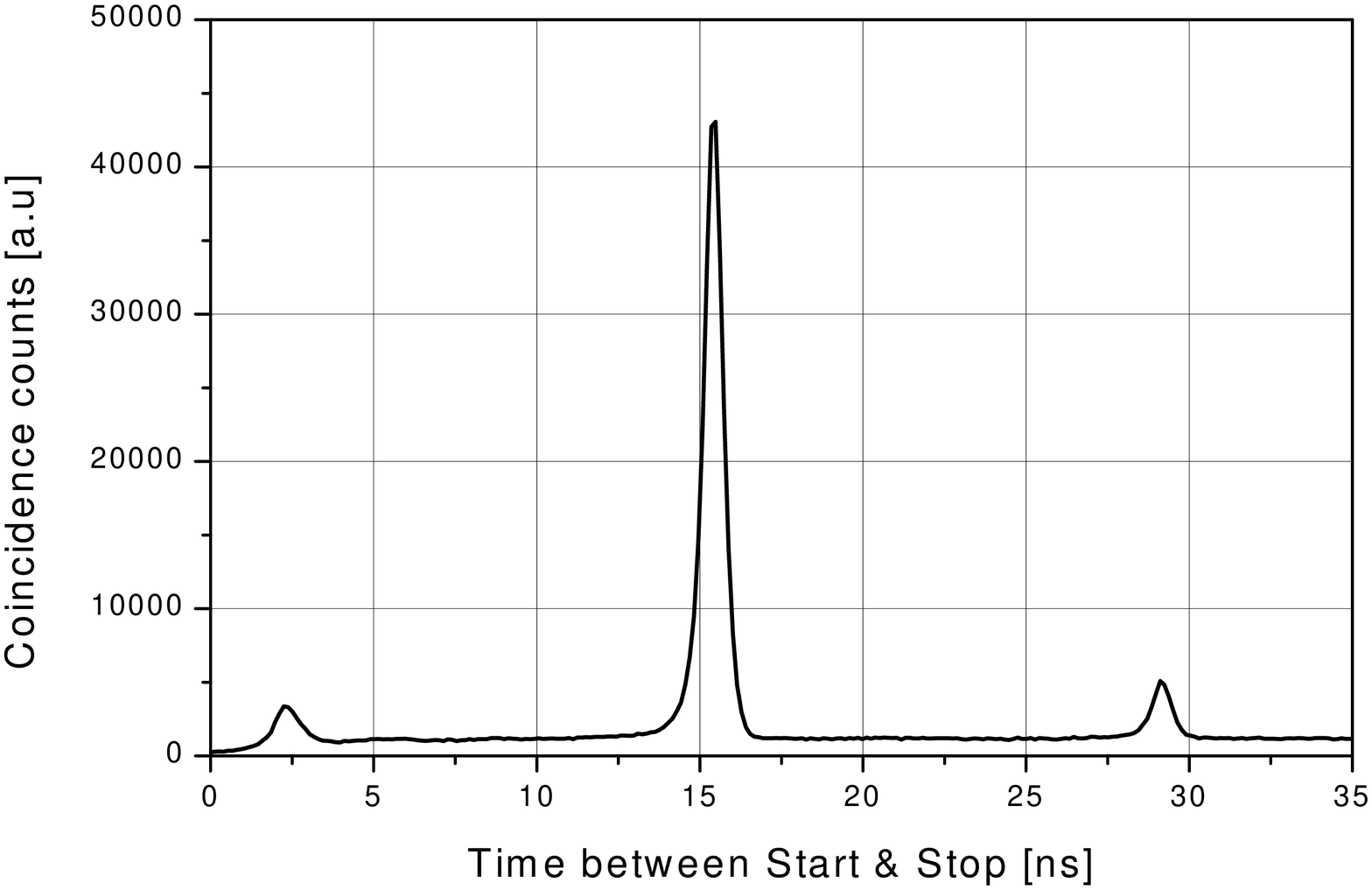}
\caption{{Time histogram of the difference of arrival times between Alice's
and Bob's detector. The spacing between two peaks is equal to the spacing
between two laser pulses}}
\label{side}
\end{figure}

We present in this section a new, easily visualized and straightforward way
of measuring this probability. The experimental setup is very simple:\newline
A series of femtosecond pulses pass through a non linear crystal creating
pairs of photons at 1310 and 1550\,nm which are separated with a WDM (see figure \ref{set side}). Each of
them is detected with the same detectors as in the previous experiment and
the difference of arrival times between Alice's and Bob's photon is measured
with a TAC.
If every created photon was detected, we would obtain only one main peak,
but because of imperfect detector efficiency, coupling and transmission losses we
observe the apparition of, what we call, side peaks (see figure \ref{side}).
These side peaks have been observed in different context as well (for
instance \cite{baby}).
The right (left) side peak is due to events where the start at Alice's side
is given by a photon created by a pulse, but where its twin is not detected
at Bob's side. The stop is then given by another photon created by the
following (preceding) pulse.  By measuring the ratio between the main peak and the side peak we obtain directly the wanted
probability :
\begin{equation}
P_{pair}=\frac{\text{counts in the side peak}}{\text{counts in the main peak}%
} \label{sideeq}
\end{equation}
This equation holds only for $t_B\eta_B\ll1$. The theoretical development is presented in the
appendix.

Figure \ref{test} depicts the pair creation rate, calculated from
the ratio of side to main peak (Eq.\ref{sideeq}), as a function of
the single count rate of the Ge-detector. The solid line shows the
prediction based on Eq.(\ref{standard}) where we estimate
$t_A=30\,\%$ and $\eta_A=9\,\%$ \cite{erreur}. We see that both
methods are in qualitative agreement, however the deviation of the
measured points from the solid line is due to the fact that in
practice $t_A\eta_A$ vary.
\begin{figure}[h]
\includegraphics[width=8.43cm]{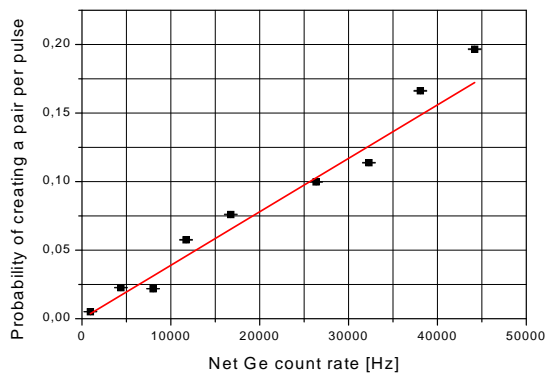}
\caption{Pair creation rate as a function of the single count
rate of the 1310 nm photon detector (hence pump power). The points are
values calculated from the ratio between side and main peaks, the
solid line is a prediction based on
Eq.\ref{standard} assuming that $t_A=30\,\%$ and $\eta_A=9\,\%$.} \label{test}
\end{figure}

Our new method has two main advantages compared to the standard one (Eq.(\ref{standard})): It is easily
visualized and it immediately
gives a good indication as to whether the probability to have
more than one pair is significant; Secondly, no estimation has to be done,
the probability is computed only from measured values and the uncertainty of,
 $P_{pair}$ is smaller than when using the method mentioned
previously (see appendix).

\section{Conclusion}

In this paper we presented a new source for realization of complex quantum
protocols over long distances. This new source is the first one creating
time-bin entangled qubits at telecom wavelengths with ultrashort pulses. We characterized this
source by performing Bell type tests, obtaining net coincidence
visibilities of up to 91 $\pm $ 0.8\,\%. We found that the change
of visibility with pump power depends on the process of creation
of the entangled time-bin qubits (femtosecond or picosecond pump pulses). Finally, we presented a new and simple
tool for measuring the probability of creating a pair per pulse.

This work was supported by the Swiss OFES in the frame of the European
QuComm IST project and by the Swiss NCCR ''Quantum Photonics''. W.T.
acknowledges funding by ESF Programme Quantum Information Theory and Quantum
Computation (QIT).

\section{Appendix}

Figure \ref{side} shows the histogram of the photons arrival time
difference at Alice's and Bob's detector. When there is a
detection in the main peak then start and stop are given by
photons created by the same pump pulse. If N is the number of
pairs created per pulse then the probability of detecting a
coincidence is given by:
\begin{equation}
{P_{\text{main peak}}=\overset{\infty }{\underset{N\geq 1}{\sum
}}}P(N\mid Start)P(Stop_{\,0}\mid N) \nonumber
\end{equation}
Here, $P(N\mid Start)$ is the probability of having N pairs
knowing that there was a start. $P(Stop_{\,0}\mid N)$ is the
probability of detecting a stop by one of the photons created by
the {\em same} pulse as the one that gave the start.

The first term can be easily computed with Bayes' rule:
\begin{equation}
P(N\mid Start)=\frac{P(N\&Start)}{P(Start)}=\frac{P(N)P(Start\mid N)}{%
P(Start)}   \nonumber
\end{equation}
where $P(N)$ is the probability that $N$ pairs are emitted. If $N$
pairs are created, the probability that the start is {\em not}
given is $(1-P(\Delta \lambda _{A}) t_{A}\eta _{A})^{N}$, where
$P(\Delta \lambda _{X})$ describes the probability that a created
photon passes through a possibly included interference filter
--- that is, $P(\Delta \lambda _{X})=1$ if there is no filter; as in the main text, $t_{X}$
characterizes the coupling ratio and transmission, and $\eta_{X}$
is the quantum efficiency of the detector. Therefore, the
probability of having a start knowing that $N$ pairs were created
is given by
\begin{equation}
P(Start \mid N)=1-(1-P(\Delta \lambda _{A}) t_{A}\eta_{A})^{N}\,.\label{5}
\nonumber
\end{equation}
Of course, $P(Start)=\overset{\infty }{\underset{M=0}{\sum}}P(M)P(Start \mid M)$,
but this is a global factor that plays
no role in what follows.

In the same way we find
\begin{eqnarray}
P(Stop_{\,0}\mid N)&=&1-\big(1-P(\Delta \lambda _{B}\mid \Delta
\lambda _{A})t_{B}\eta _{B}\big)^{N}\nonumber
\end{eqnarray}
where $P(\Delta \lambda _{B}\mid \Delta \lambda _{A})$ is the
probability that a photon at Bob's side passes through an
interference filter knowing that its twin photon has already
passed through an interference filter at Alice's side, thus
$P(\Delta \lambda _{B}\mid \Delta \lambda _{A})=1$ when $ \Delta
\lambda _{B}\geq \Delta \lambda_{A}$ \cite{foot21}. We assume that
the spectrum of the created photons is centered at the maximum
transmission of the interference filters.

The probability of detecting a coincidence in the right side peak
is given by:
\begin{multline}
P_{\text{side peak}}=\\
\overset{\infty }{\underset{N\geq 1}{\sum }}P(N\mid
Start)\,(1-P(Stop_{\,0}\mid N))\,P(Stop_{\,1})  \label{11}
\nonumber
\end{multline}
The first term represents, as before, the probability of having N
pairs knowing that there was a start, the second is the
probability not to detect a stop originating from the same pump
pulse; $P(Stop_{\,1})$ is the probability that the stop is given by
a photon created by the first pulse following the one which gave
the start. Explicitly
\begin{equation}
P(Stop_{\,1}) =\overset{\infty }{\underset{M=0}{\sum}}P(M)\,\big[1-(1-P(\Delta \lambda _{B}) t_{B}\eta_{B})^{M}\big]
\nonumber
\end{equation}
note that here we have $P(\Delta \lambda _{B})$ instead of
\newline
$P(\Delta \lambda _{B}\mid \Delta \lambda _{A})$, since we don't
require that the twin photon has passed through the corresponding
filter.

We now suppose that the mean number of pairs is much smaller than
1, so that $P(N>1)=0$ and $P(1)=P_{pair}$. From the equations
above, we find the ratio between main and side peak to be:
\begin{multline}
\frac{P_{\text{main peak}}}{P_{\text{side peak}}}= \\
\frac{P(\Delta \lambda _{B}\mid \Delta \lambda _{A})}{%
P_{pair}(1-P(\Delta \lambda _{B}\mid \Delta \lambda _{A})t_{B}\eta
_{B})P(\Delta \lambda _{B})} \label{16} \nonumber
\end{multline}
If there is only a filter at Alice's side (as was in our Bell type
experiment) and $t_B\eta_B\ll 1$, we find equation (\ref{sideeq})
($P(\Delta \lambda _{B}\mid \Delta \lambda _{A})=1$ and $P(\Delta
\lambda _{B})=1$). Thus, if one wants to measure the probability
of creating a pair per pulse in a given spectral bandwidth, one
has to filter both photons.
\newline Finally, using this method, the uncertainty of $P_{pair}$ is
reduced compared to the standard method (Eq.\ref{standard}). For
instance, if we estimate $t_{B}=30\pm6\,\%$ and $\eta _{B}=30\pm
6\,\%$, then the relative uncertainty of $P_{pair}$, calculated
using Eq.\ref {standard}, is 30\,\% while it is only of 3\,\%
using
our new method (Eq.\ref{sideeq}).


\begin{thebibliography}{99}
\bibitem{foot2}  The main reason why we use a bulk interferometer is to avoid
high chromatic dispersion at 710\,nm in optical fibers
\bibitem{foot21}  Note that the spectral filtering of one entangled photon also
filters its twin photon
\bibitem{foot3}  To affirm that we violate Bell inequalities, we have to
make the (reasonable) assumption that the other (not recorded) triple
coincidence have the same behavior and that all coincidence count rates
depend only on the sum of the phases
\bibitem{erreur}  The estimation of Alice's coupling ratio and transmission ($t_A$)
is made by assuming that it is comparable to Bob's ($t_B$) which
can be measured knowing the coincidence count rate, the count rate
of the Ge-detector and the quantum efficiency of InGaAs APD.
\end{thebibliography}
\end{document}